\documentclass[11pt,a4paper,twoside]{article}
\usepackage{amsmath}
\usepackage{amsfonts}
\usepackage{changes}
\numberwithin{equation}{section}

\begin{document}

\noindent

{\bf
{\Large Massive gravity and Fierz-Pauli theory}
} 

\vspace{.5cm}
\hrule

\vspace{1cm}

\noindent

{\large\bf{Alberto Blasi\footnote{\tt alberto.blasi@ge.infn.it }
and Nicola Maggiore\footnote{\tt nicola.maggiore@ge.infn.it }
\\[1cm]}}

\setcounter{footnote}{0}

\noindent
%{\small{
{{}Dipartimento di Fisica, Universit\`a di Genova,\\
via Dodecaneso 33, I-16146, Genova, Italy\\
and\\
{} I.N.F.N. - Sezione di Genova\\
%}}
\vspace{1cm}

\noindent
{\tt Abstract~:}
Linearized gravity is considered as an ordinary gauge field theory. This implies the need for gauge fixing in order to have well defined propagators. Only after having achieved this, the most general mass term is added. The aim of this paper is to study of the degrees of freedom of the gauge fixed theory of linearized gravity with mass term. The main result  is that, even outside the usual Fierz-Pauli constraint on the mass term, it is possible to choose a gauge fixing belonging to the Landau class, which leads to a massive theory of gravity with the five degrees of freedom of a spin two massive particle. 

\newpage

%**************************************************************************
\section{Introduction}
%**************************************************************************

In this paper we consider the theory of linearized gravity with mass term. In particular, we are concerned with the well known problem of the counting of the degrees of freedom (dof). We will not list here the many good reasons to modify gravity at long distances, referring the reader to the many and thorough reviews on this subject (see, for instance \cite{Rubakov:2008nh,Hinterbichler:2011tt,deRham:2014zqa,Blake:2013bqa,Amoretti:2017xto,Amoretti:2014zha,Amoretti:2014mma,Vegh:2013sk}). Our starting point is the assumption that the need for a massive gravity does exist. The graviton of linearized gravity is represented by $h_{\mu\nu}(x)$, which is a rank-2 symmetric tensor field. Hence, in a $D$-dimensional spacetime, the number of independent components is $D(D+1)/2$. If $D=4$, the number of components is 10. In $4D$ massless general relativity, the physical dof are reduced by four constraints and four general coordinate transformations, leading to the two degrees of freedom of a massless spin two graviton. The mass term breaks the four diffeomorphism symmetries, leaving the four constraints, so that the number of independent components seems to be six. On the other hand, we know that the dof of a massive spin-2 particle should be five, the sixth being a ghost (the so called Boulware-Deser ghost \cite{Boulware:1973my}). It is an accepted fact that, if the mass term is constrained to be a particular one, $i.e.$ the Fierz-Pauli (FP) term \cite{Fierz:1939ix}, the sixth ghost dof is removed. There is, nonetheless, a price to pay. Indeed, if on one hand the FP model  displays the correct number of dof, on the other it suffers from several drawbacks, like, for instance, the intrinsic impossibility of the existence of a good massless limit \cite{Boulware:1973my}. 

	The motivation of this work comes from the observation that, from a pure field theoretical point of view the theory of linearized gravity is a gauge field theory, invariant under a gauge transformation of the field $h_{\mu\nu}(x)$. Hence,  the first concern should be to add an invariant gauge fixing term, in order to have a well defined theory with well defined propagators (see, for instance \cite{Baulieu:1983tg}. Only after having done this, one might face the problem of adding a mass term, for whatever reason it turns out necessary or convenient to do this. In general, the mass term breaks gauge invariance, and field theory provides us various ways to limit the damage. But one thing is to add a mass term to a well defined theory, and then trying to control the breaking, like it is done, for instance, in supersymmetric gauge field theories \cite{Girardello:1981wz,Maggiore:1996gg}. A different one is to add directly a mass term to an ill defined theory, where the propagators are not defined $ab\ initio$ because of the presence of a  gauge symmetry, and then realizing that the resulting theory is affected by several problems. This is the result of the double use of a term which is at the same time treated as a mass term and as gauge fixing term: without the mass term  the invariant action does not have propagators at all. The deal of proceeding that way is that the FP action has the correct number of dof, without BD ghost, at the additional price of having an highly constrained mass term with one free parameter only instead of the original two, because of the FP constraint on the masses.

The purpose of our research is trying to kill two birds with one stone, $i.e.$ 
\begin{enumerate}
\item constructing a well defined, massless,  field theory for the spin-2 graviton $h_{\mu\nu}(x)$, invariant under the gauge symmetry, with well defined propagators by means of an invariant gauge fixing term
\item adding a mass term, which breaks gauge invariance and which modifies also the propagators, hence asking which are the constraints on the parameters  in order that the massive, symmetry broken theory has well defined propagators, namely without tachyonic unphysical poles induced by the mass term
\item addressing, finally, the question of the five dof, $i.e.$ selecting amongst the solutions satisfying the previous requirements 1. and 2. , those, if any, which display the correct number of dof, in the hope of going beyond FP theory, finding a model with a more general, less constrained  mass term which nonetheless displays five dof.
\end{enumerate}

The first two points were solved in a previous paper \cite{Blasi:2015lrg}. In the present work we address the  physically crucial third point, concerning the dof. 

The paper is organized as follows. In Section 2 we recall the FP theory, with the aim of testing the procedure we follow to study the dof of the theory. We recover, of course, the five dof of a massive spin-2 field. In Section 3 we briefly summarize the results of our previous work \cite{Blasi:2015lrg}, listing the (eight) possibilities for a theory of a massive symmetric rank-2 two tensor field with well defined propagators, free of unphysical tachyonic poles. Lacking an analysis of the dof, these models are not theories of massive gravity yet, since only a massive rank-2 tensor field with five dof has the right of being called graviton. In Section 4 the analysis of the dof is done, using the technique presented in Section 2. In Section 5, we summarize and discuss our results.

%**************************************************************************
\section{The Fierz-Pauli case}
%**************************************************************************

The action of linearized gravity in $D$ spacetime dimensions is given by 
\begin{equation}
S_{inv}=\int d^Dx
\left(
\frac{1}{2}h\partial^2h - h_{\mu\nu}\partial^\mu\partial^\nu h - \frac{1}{2}h^{\mu\nu}\partial^2h_{\mu\nu} + h^{\mu\nu}\partial_\nu\partial^\rho h_{\mu\rho}
\right),
\label{2.1}\end{equation}
which represents the most general 4D action written in terms of a rank-2 symmetric tensor $h_{\mu\nu}(x)=h_{\nu\mu}(x)$ invariant under the gauge transformation
\begin{equation}
\delta h_{\mu\nu}=\partial_\mu\theta_\nu + \partial_\nu\theta_\mu \label{2.2}.
\end{equation}
In \eqref{2.1}, $h(x)$ is the trace of $h_{\mu\nu}(x)$: $h(x)=h^\lambda_{\ \lambda}(x)=\eta^{\mu\nu}h_{\mu\nu}(x)$, where $\eta_{\mu\nu}=\mbox{Diag}(-1,1,\ldots,1)$.

The need to modify General Relativity at large distances is the main motivation for the massive deformation of the linearized theory of gravity  by the following mass term 
\begin{equation}
S_m=\int d^Dx \frac{1}{2} \left (
m_1^2h^{\mu\nu}h_{\mu\nu} + m_2^2 h^2
\right),\label{2.3}
\end{equation}
where $m_1^2$  and $m_2^2$ are  mass parameters for $h_{\mu\nu}(x)$ and $h(x)$, respectively. 

The mass term breaks the gauge invariance \eqref{2.2}, and it is constrained by the counting of the degrees of freedom (dof). In fact, in absence of mass term, in $D=4$ spacetime dimensions (4D) gravitational waves described by $h_{\mu\nu}(x)$ are characterized by two propagating dof: the two physical modes of a massless graviton\cite{Carroll:2004st}. In presence of a generic mass term like \eqref{2.3}, general covariance is broken and the dof of the theory described by the action $S_{inv}+S_m$  are 6. On the other hand, we know that  the theory of a massive spin $S=2$ tensor field in 4D must have $2S+1=5$ dof.  The sixth dof is the so-called Boulware-Deser (BD) scalar ghost \cite{Boulware:1973my}. Removing the sixth ``bad'' dof is uniquely realized by the following choice of the mass term $S_m$ \eqref{2.3}: 
\begin{equation}
m_1^2+m_2^2=0,
\label{2.4}\end{equation}
which is called the Fierz-Pauli (FP) choice \cite{Fierz:1939ix}.

We show now that, indeed, if the action is given by 
\begin{equation}
S_{FP}=S_{inv}+S_m,
\label{2.5}\end{equation} 
the FP choice \eqref{2.4} is the only one which yields a massive theory of gravity describing only five propagating dof, without BD ghost.

In momentum space, 
%($\partial_\mu=ip_\mu$)
the equations of motion (eom) deriving from $S_{FP}$ \eqref{2.5} for  $\tilde{h}_{\mu\nu}(x)$, which is the Fourier transform of $h_{\mu\nu}(x)$, are
\begin{eqnarray}
\frac{\delta S_{FP}}{\delta\tilde{h}_{\mu\nu}} 
&=&
[(p^2-m_2^2)\eta^{\mu\nu}-p^\mu p^\nu]\tilde{h} - 
(p^2+m_1^2)\tilde{h}_{\mu\nu} 
\nonumber \\ &&
+
(p^\mu p_\lambda\tilde{h}^{\nu\lambda}+p^\nu p_\lambda\tilde{h}^{\mu\lambda}) -
\eta^{\mu\nu}p^\lambda p^\tau \tilde{h}_{\lambda\tau} 
=0.
\label{2.6}
\end{eqnarray}
Let us introduce the projection operator
\begin{equation}
%d_{\mu\nu} \equiv  \eta_{\mu\nu}- e_{\mu\nu}\ ;\
e_{\mu\nu} \equiv \frac{p_\mu p_\nu}{p^2}, 
\label{2.7}\end{equation}
and saturate \eqref{2.6} with $\eta_{\mu\nu}$, $e_{\mu\nu}$ and $p_\mu$ . We get, respectively
\begin{eqnarray}
\eta_{\mu\nu}\frac{\delta S_{FP}}{\delta\tilde{h}_{\mu\nu}} 
&=&
[p^2(D-2)-(m_1^2+Dm_2^2)]\tilde{h} 
+p^2e_{\mu\nu}\tilde{h}^{\mu\nu}(2-D) = 0 \label{2.8} \\
e_{\mu\nu}\frac{\delta S_{FP}}{\delta\tilde{h}_{\mu\nu}}
&=&
m_2^2\tilde{h}+m_1^2e_{\mu\nu}\tilde{h}^{\mu\nu} = 0 \label{2.9}\\
p_\mu\frac{\delta S_{FP}}{\delta\tilde{h}_{\mu\nu}}
&=&
m_1^2p_\mu\tilde{h}^{\mu\nu}+m_2^2p^\nu\tilde{h} = 0. \label{2.10}
\end{eqnarray}
Supposing $m_1^2\neq 0$, from \eqref{2.9} we find
\begin{equation}
e_{\mu\nu}\tilde{h}^{\mu\nu}=-\frac{m_2^2}{m_1^2}\tilde{h},
\label{2.11}\end{equation}
so that \eqref{2.8} becomes
\begin{equation}
(D-2)\frac{m_1^2+m_2^2}{m_1^2}p^2\tilde{h} - (m_1^2+Dm_2^2)\tilde{h} =0,
\label{2.12}\end{equation}
and with the FP condition \eqref{2.4} the coefficient of $p^2$ vanishes. We thus have
\begin{equation}
\tilde{h}=0,
\label{2.13}\end{equation}
and, from \eqref{2.10}
\begin{equation}
p_\mu\tilde{h}^{\mu\nu}=0.
\label{2.14}\end{equation}
Hence, the FP condition $m_1^2+m_2^2=0$ implies \eqref{2.13} and \eqref{2.14}, which represent the $1+D$ constraints which, subtracted from the $D(D+1)/2$ components of the D-dimensional rank-2 symmetric tensor field $h_{\mu\nu}(x)$ leave us with the correct $2S+1$ dof of a massive spin $S=2$ field in $D$ spacetime dimensions:
\begin{equation}
\left.\underbrace{\frac{D(D+1)}{2}}_{\text{$h_{\mu\nu}=h_{\nu\mu}$}}
-\underbrace{1}_{\text{$h=0$}}
-\underbrace{D}_{\text{$\partial_\mu h^{\mu\nu}=0$}}
=
2S+1\right|_{D=4;S=2}=5.
\label{2.15}\end{equation}

%**************************************************************************
\section{The gauge fixed model}
%**************************************************************************

In order to have a well defined gauge field theory, according to the usual rules of quantum field theory \cite{Baulieu:1983tg}, we first add to the invariant action $S_{inv}$ \eqref{2.1} the following gauge fixing term
\begin{equation}
S_{gf} = \int d^4x \left[
b^\mu (\partial^\nu h_{\mu\nu} +\kappa_1\partial_\mu h) +\frac{\kappa}{2}b^\mu b_\mu \right]\label{3.1},
\end{equation}
which, by means of a Nakanishi-Lautrup Lagrange multiplier $b_\mu(x)$ \cite{Nakanishi:1966zz,Lautrup:1967zz}, implements the gauge condition
\begin{equation}
\partial^\nu h_{\mu\nu} +\kappa_1\partial_\mu h=0.\label{3.2}
\end{equation}
From \eqref{3.1} and \eqref{3.2} we see that the gauge fixing procedure for the symmetric tensor field $h_{\mu\nu}(x)$ implies the presence of two gauge coupling constants: $\kappa$  and $\kappa_1$. The gauge fixed action $S_{inv}+S_{gf}$ given by the sum of \eqref{2.1} and \eqref{3.1} has well defined propagators and is invariant under a BRS symmetry \cite{Becchi:1974xu}. To this action we add the mass term \eqref{2.3}, so that the total action we are considering for massive gravity is given by
\begin{equation}
S=S_{inv}+S_{gf}+S_m.
\label{3.3}\end{equation}

The theory defined by \eqref{3.3} depends on four parameters: $m^2_1, \ m_2^2, \ \kappa$ and $\kappa_1$. In \cite{Blasi:2015lrg}, we determined which are the constraints on them in order to have a theory without tachyonic poles in the propagators. We found eight possible solutions: 
\begin{description}
\item[solution 1] %no FP
\begin{equation}
m_1^2> 0 \ ;\ m_1^2+Dm_2^2= 0\ ;\ \kappa<0\ ;\ \kappa_1=0.
\label{3.4}\end{equation}

\item[solution 2] %FP ok
\begin{equation}
m_1^2>0\ ;\ 
m_1^2+Dm_2^2<0\ ;\ 
\kappa\leq\frac{D-1}{2-D}\ ;\ \kappa_1=-1
\label{3.5}\end{equation}
\item[solution 3] %no FP
\begin{equation}
m_1^2>0\ ;\ 
m_1^2+Dm_2^2>0\ ;\ 
\frac{D-1}{2-D}\leq\kappa<0\ ;\ \kappa_1=-1
\label{3.6}\end{equation}
\item[solution 4] %no FP
\begin{equation}
m_1^2=0\ ;\
m_2^2<0\ ;\ \kappa\leq\frac{D-1}{2-D}\ ;\ \kappa_1=-1
\label{3.7}\end{equation}
\item[solution 5] %no FP
\begin{equation}
m_1^2=0\ ;\
m_2^2>0\ ;\ 
\frac{D-1}{2-D}\leq\kappa<0\ ;\ \kappa_1=-1
\label{3.8}\end{equation}
\item[solution 6] %no FP
\begin{equation}
m_1^2=0\ ;\
m_2^2<0\ ;\ \kappa>0\ ;\ \kappa_1=-1
\label{3.9}\end{equation}

\item[solution 7] %FP ok
\begin{equation}
m_1^2 \geq 0 \ ;\ m_1^2+Dm_2^2 \leq 0\ ;\ \kappa=0\ ;\ \kappa_1=0
\label{3.10}\end{equation}

\item[solution 8] %FP ok
\begin{equation}
m_1^2\geq 0 \ ;\ m_1^2+Dm_2^2\neq 0\ ;\ \ ;\ \kappa=0\ ;\ \kappa_1=-1
\label{3.11}\end{equation}
\end{description}
All and only the above solutions display well defined propagators, hence they represent good field theories for a symmetric rank-2 tensor field, characterized by the action $S_{inv}$ \eqref{2.1}, invariant under the gauge symmetry \eqref{2.2}, which is gauge fixed by $S_{gf}$ \eqref{3.1}, and is broken by the most general mass term $S_m$ \eqref{2.3}. Concerning the mass term, five out of eight solutions, in particular solutions n. 1, 3, 4, 5 and 6,  exclude the FP point \eqref{2.4}, and this, of course, begs the question of which, amongst the eight solutions above, represent $also$ a theory of massive gravity, $i.e.$ a theory for a symmetric rank-2 tensor field $h_{\mu\nu}(x)$ characterized by {\it five dof only}.

%**************************************************************************
\section{Degrees of freedom}
%**************************************************************************

In order to determine the dof of each of the eight candidates, we proceed like we did for the pure FP theory in Section 2. Our aim is to investigate whether from the eom of the full action $S$ \eqref{3.3} it is possible to extract the two conditions (in momentum space)
\begin{eqnarray}
p_\nu\tilde{h}^{\mu\nu} &=& 0 \label{4.1} \\
\tilde{h} &=& 0 \label{4.2}
\end{eqnarray}
which allow for the counting \eqref{2.15} which leads to 5 dof in D=4 spacetime dimensions.

The eom for the action $S$ \eqref{3.3} are
\begin{eqnarray}
\frac{\delta S}{\delta\tilde{h}_{\mu\nu}}
&=&
\frac{\delta S_{FP}}{\delta\tilde{h}_{\mu\nu}} + \frac{\delta S_{m}}{\delta\tilde{h}_{\mu\nu}} \nonumber \\
&=&
\frac{\delta S_{FP}}{\delta\tilde{h}_{\mu\nu}} 
-\frac{i}{2}(p^\mu\tilde{b}^\nu+p^\nu\tilde{b}^\mu)
-i\kappa_1\eta^{\mu\nu}p_\lambda\tilde{b}^\lambda
=0 \label{4.3}\\
\frac{\delta S}{\delta\tilde{b}_{\mu}} 
&=&
\kappa\tilde{b}^\mu + ip_\nu\tilde{h}^{\mu\nu} + i\kappa_1p^\mu\tilde{h} = 0, \label{4.4}
\end{eqnarray}
where $\frac{\delta S_{FP}}{\delta\tilde{h}_{\mu\nu}}$ is given by \eqref{2.6} and $\tilde{b}_{\mu}(x)$ is the Fourier transform of ${b}_{\mu}(x)$.

Saturating \eqref{4.3} with $\eta_{\mu\nu}$ and $e_{\mu\nu}$, like we did in the pure FP case,  we get
\begin{eqnarray}
\eta_{\mu\nu}\frac{\delta S_{FP}}{\delta\tilde{h}_{\mu\nu}} 
&=& 
i(1+D\kappa_1)p_\lambda\tilde{b}^\lambda \label{4.5} \\
e_{\mu\nu}\frac{\delta S_{FP}}{\delta\tilde{h}_{\mu\nu}}
&=&  
-i(1+\kappa_1)p_\lambda\tilde{b}^\lambda \label{4.6}
%\\
%p_\mu\frac{\delta S_{FP}}{\delta\tilde{h}_{\mu\nu}} 
%&=& 
%-\frac{i}{2}p^2\tilde{b}^\nu - i(\kappa_1+\frac{1}{2})p^\nu p_\lambda\tilde{b}^\lambda, \label{4.7}
\end{eqnarray}
which generalize \eqref{2.8} and \eqref{2.9} 
%and \eqref{2.10} 
to the gauge fixed case.
%Moreover, saturating \eqref{4.4} with $p_\mu$ we have
%\begin{equation}
%p_\mu\frac{\delta S}{\delta\tilde{b}_{\mu}} 
%=
%\kappa p_\lambda\tilde{b}^\lambda + i p^2 (e_{\mu\nu}\tilde{h}^{\mu\nu} + \kappa_1 \tilde{h}) = 0.
%\label{4.8}\end{equation}

The idea is to fit the eight solutions \eqref{3.4}-\eqref{3.11} to the conditions \eqref{4.5} and \eqref{4.6}  to see if any of them leads to the constraints \eqref{4.1} and \eqref{4.2} needed in order to have the desired five dof of massive gravity.

Let us illustrate the procedure for the solution n. 2 \eqref{3.5}, which can be taken as an example of what typically happens. Notice that this is one of the solutions which include the FP point \eqref{2.4}, for which we know that five dof are realized. Putting the values for the parameters $m_1^2$, $m_2^2$, $\kappa$ and $\kappa_1$  of solution 2, the equations \eqref{4.4}, \eqref{4.5} and \eqref{4.6} 
%and \eqref{4.7} 
become, respectively 
\begin{equation}
\kappa\tilde{b}^\mu + ip_\nu\tilde{h}^{\mu\nu} - ip^\mu\tilde{h} = 0 
\label{4.7}\end{equation}
\begin{equation}
[p^2(D-2)-(m_1^2+Dm_2^2)]\tilde{h} + (2-D) p^2e_{\mu\nu}\tilde{h}^{\mu\nu} 
= i(1-D)p_\mu\tilde{b}^\mu 
\label{4.8} \end{equation}
\begin{equation}
m_2^2\tilde{h}+m_1^2e_{\mu\nu}\tilde{h}^{\mu\nu} = 0 
\label {4.9}\end{equation}
%\begin{equation}
%m_1^2p_\mu\tilde{h}^{\mu\nu}+m_2^2p^\nu\tilde{h} = \frac{i}{2}(-p^2\tilde{b}^\nu+p^\nu p_\mu\tilde{b}^\mu). \label{4.11} 
%\end{equation}
Supposing $m_2^2\neq 0$\footnote{The cases $m_1^2,m_2^2=0$ do not lead to any identification of the dof of this solution}, from \eqref{4.9} we get
\begin{equation}
\tilde{h}=-\frac{m_1^2}{m_2^2}e_{\mu\nu}\tilde{h}^{\mu\nu}
\label{4.10}\end{equation}
hence, from \eqref{4.7}
\begin{equation}
\tilde{b}^\mu=-\frac{i}{\kappa}\frac{m_1^2+m_2^2}{m_2^2}p_\nu\tilde{h}^{\mu\nu}.
\label{4.11}\end{equation}
Substituting into \eqref{4.8}, we have
\begin{equation}
(m_1^2+m_2^2)(D-2+\frac{1-D}{\kappa})p^2e_{\mu\nu}\tilde{h}^{\mu\nu}=(m_1^2+Dm_2^2)e_{\mu\nu}\tilde{h}^{\mu\nu}
\label{4.12}\end{equation}
Notice that for the solution 2 \eqref{3.5}, we have that $D-2+\frac{1-D}{\kappa}\neq0$ and $m_1^2+Dm_2^2\neq0$, hence if the FP condition \eqref{2.4} is satisfied $m_1^2+m_2^2=0$, from \eqref{4.12} and \eqref{4.11} we get, respectively 
\begin{equation}
e_{\mu\nu}\tilde{h}^{\mu\nu} =\tilde{b}^\mu= 0
\label{4.13}\end{equation}
so that, from \eqref{4.10} 
\begin{equation}
\tilde{h}=0
\label{4.14}\end{equation}
and, from \eqref{4.8}
\begin{equation}
p_\nu\tilde{h}^{\mu\nu}.
\label{4.15}\end{equation}
The conditions \eqref{4.14} and \eqref{4.15} are those  needed to have 5 dof , according to \eqref{2.15}. Therefore,  solution 2 represents 5 dof only if collapses to the FP point. Elsewhere in the generic range of masses concerned by this solution ($m_1^2+Dm_2^2<0$) , no easy identification of the dof of the theory can be obtained. Still, an advantage exists in comparison to the pure FP theory described in Section 2  : solution 2  is an example of a well defined gauge field theory, with no tachyonic poles in the propagators, for a symmetric rank-2 tensor field, where the 5 dof 
are realized at the FP point \eqref{2.4}, hence it represents a theory of massive gravity.

Solution 2 is an example of what typically happens for all the eight solutions displayed in Section 3: outside the FP point, we are not able to get $both$ the conditions \eqref{2.13} and \eqref{2.14}  which realize the schema \eqref{2.15} leading to 5 dof: either none of them, or only one of them. But an exception does exist, represented by solution 8 \eqref{3.11} ($m_1^2\geq 0 \ ;\ m_1^2+Dm_2^2\neq 0\ ;\ \ ;\ \kappa=0\ ;\ \kappa_1=-1$), which includes the FP point \eqref{2.4}.  

The general equations \eqref{4.4}, and \eqref{4.6}, written for the solution 8,  are: 
\begin{equation}
p_\nu\tilde{h}^{\mu\nu} - p^\mu\tilde{h} = 0 
\label{4.16}\end{equation}
%\begin{equation}
%[p^2(D-2)-(m_1^2+Dm_2^2)]\tilde{h} + (2-D) p^2e_{\mu\nu}\tilde{h}^{\mu\nu} = i(1-D)p_\mu\tilde{b}^\mu \label{4.19} 
%\end{equation}
\begin{equation}
m_2^2\tilde{h}+m_1^2e_{\mu\nu}\tilde{h}^{\mu\nu} = 0.
\label {4.17}\end{equation}
%\begin{equation}
%m_1^2p_\mu\tilde{h}^{\mu\nu}+m_2^2p^\nu\tilde{h} = \frac{i}{2}(-p^2\tilde{b}^\nu+p^\nu p_\mu\tilde{b}^\mu). \label{4.21} 
%\end{equation}

%**************************
%
%Saturating \eqref{4.16} with $p_\mu$, we get
%\begin{equation}
%p^2e_{\mu\nu}\tilde{h}^{\mu\nu}=p^2\tilde{h}.
%\label{4.18bis}\end{equation} 
%
%From \eqref{4.17}, {\it provided that $m_1^2\neq0$}, we have
%\begin{equation}
%e_{\mu\nu}\tilde{h}^{\mu\nu}=-\frac{m_2^2}{m_1^2}\tilde{h}\Rightarrow
%p^2e_{\mu\nu}\tilde{h}^{\mu\nu}=-\frac{m_2^2}{m_1^2}p^2\tilde{h}.
%\label{4.19bis} \end{equation}
%The equations \eqref{4.18} and \eqref{4.19} are compatible only if $-\frac{m_2^2}{m_1^2}=1$, $i.e.$ on the FP point.
%
%If, on the other hand, we consider the case $m_1^2=0$,   the equation \eqref{4.17} then implies
%\begin{equation}
%\tilde{h}=0,
%\label{4.20bis}\end{equation}
%which, substituted in \eqref{4.16} gives
%\begin{equation}
%p_\nu\tilde{h}^{\mu\nu}=0.
%\label{4.21bis}\end{equation}
%Hence, quite remarkably, we get the two conditions needed for the realization of the 5 dof of a massive spin 2 massive particle also outside the FP point. 
%
%**************************

We can write
\begin{equation}
e_{\mu\nu}\tilde{h}^{\mu\nu}=\frac{p_\mu}{p^2}p_\nu\tilde{h}^{\mu\nu}=\tilde{h},
\label{4.18}\end{equation} 
where we used \eqref{4.16} in the last equality.
Using this, from \eqref{4.17} we get 
\begin{equation}
(m_1^2+m_2^2)\tilde{h}=0.
\label{4.19}\end{equation}
This condition is satisfied either on the FP point $m_1^2+m_2^2=0$, or else outside the FP point, provided that 
\begin{equation}
\tilde{h}=0,
\label{4.20}\end{equation}
which, substituted into \eqref{4.16}, gives 
\begin{equation}
p_\nu\tilde{h}^{\mu\nu}=0.
\label{4.21}\end{equation}

Hence, quite remarkably, we get the two conditions needed for the realization of the 5 dof of a massive spin 2 massive particle also outside the FP point (provided that $m_1^2+Dm_2^2\neq 0$). 

%*****************************************************************}
\section{Summary and conclusions}
%*****************************************************************}

In this paper we considered the theory of linearized gravity, defined by the action \eqref{2.1} written in terms of a rank-2 symmetric tensor field $h_{\mu\nu}(x)$, which is commonly called ``graviton''. The theory is taken as an ordinary gauge field theory, defined by the gauge transformation \eqref{2.2}, which indeed uniquely identifies the action \eqref{2.1} as the most general invariant action. Under this respect, the analogy with the electromagnetism is complete: the action \eqref{2.1} stands to the Maxwell action as the graviton $h_{\mu\nu}(x)$ stands to the photon $A_\mu(x)$. This field theoretical approach implies that the first task should be the definition of a generating functional of the Green functions, the simplest of which are  the 2 points Green functions, $i.e.$ the propagators, which indeed are not defined for the action \eqref{2.1} alone. Like in electromagnetism, a gauge fixing is needed. The most general gauge fixing term for a rank-2 tensor field is given by \eqref{3.1}, which corresponds to the gauge condition \eqref{3.2}, which generalizes to a rank-2 tensor field the gauge condition $\partial^\mu A_\mu(x)$ for the electromagnetic vector field. Consequently, the gauge field theory for $h_{\mu\nu}(x)$ depends on two gauge parameters ($\kappa$ and $\kappa_1$), instead of one like in Maxwell theory. To this general class of gauge fixing belong all the gauge fixing conditions used for linearized gravity, like for instance the harmonic gauge \cite{Carroll:2004st}. Pursuing the analogy, $\kappa=0$ defines the Landau gauge, which in the gravity side should be called Landau $gauges$, since $\kappa_1$ can still assume any value. After the gauge fixing procedure, the theory described by the invariant action \eqref{2.1} plus the gauge fixing term \eqref{3.1} is a well defined gauge field theory for the (massless) graviton $h_{\mu\nu}(x)$. 

The most general mass term for a rank-2 tensor field is given by \eqref{2.3}, and depends on two mass parameters, one for $h_{\mu\nu}(x)$ and the other for its trace $h(x)$. The introduction of the mass has two kind of consequences. The first is to break explicitly the gauge invariance \eqref{2.2}. The second one, less evident, is the introduction of unphysical tachyonic poles in the propagators. In a previous paper \cite{Blasi:2015lrg} we faced this latter problem, tuning the four parameters of the theory -two masses, two gauge parameters- in order to get rid of the bad propagators. This selected eight possibilities, listed in this paper from \eqref{3.4} to \eqref{3.11}, which identify eight classes of theories for a massive, rank-2 symmetric tensor field $h_{\mu\nu}(x)$ with well defined propagators, free of unphysical tachyons. In the massive case, we cannot call anymore $h_{\mu\nu}(x)$ ``graviton'', because of the counting of the dof: we are allowed to identify $h_{\mu\nu}(x)$ as a massive graviton only for those theories, amongst our eight candidates, which display five dof, because five are the dof a massive spin-2 field.  

It is not trivial at all to realize five dof in massive gravity, since simply breaking the gauge (or infinitesimal diffeomorphism) invariance by means of an unconstrained mass term leads, in four spacetime dimensions, from ten  to six dof, the extra dof being the Boulware-Deser ghost \cite{Boulware:1973my}. The traditional, generally accepted way out, is to constrain the mass term \eqref{2.3} with the condition on the masses \eqref{2.4}, which leads to the well known Fierz-Pauli theory \cite{Fierz:1939ix}. 

If on one hand, the Fierz-Pauli theory realizes the desired five dof of the massive graviton, on the other is not satisfactory form the field theory point of view: we would have not proceeded that way in the (quite analogous) case of electromagnetism. Adding a mass term directly to a gauge invariant action, skipping the ordinary gauge fixing procedure, seems to be a bit rude: the mass term, which breaks the gauge invariance, is used {\it at the same time} as a gauge fixing  (without it the propagators are not defined) and as a mass term. The parameters appearing in the mass term are {\it at the same time} masses (hence hopefully physical) and gauge parameters (definitely unphysical). 

The main motivation of this work was to check whether, following the standard field theoretical procedure (invariant action $\rightarrow$ gauge fixing $\rightarrow$ mass term) instead of the more common one (invariant action $\rightarrow$ mass term $\equiv$ gauge fixing), the five dof of a massive spin-2 graviton could be obtained.

Five dof are obtained out of the ten components of the symmetric rank-2 tensor field $h_{\mu\nu}(x)$ if the two conditions  $h(x)=0$ \eqref{2.13} and $\partial_\nu h^{\mu\nu}(x)=0$ \eqref{2.14} hold. In field theory, any condition should not be put by hand, but should be a consequence of some constraints on the theory, like for instance the equations of motion, carefully and conveniently managed. To reach our aim, we adopted a procedure which we first tested in Section 2 on the standard Fierz-Pauli theory, getting the known result. In Section 4 we treated in the same way the cases listed in Section 3, characterized by having good propagators. 

The  outcome is that for seven out of eight cases, either the two conditions are not obtained, or are indeed recovered, but only if the Fierz-Pauli condition \eqref{2.4} on the masses is satisfied. Remarkably, we found that the solution 8 \eqref{3.11}, which corresponds to the particular Landau gauge condition $\kappa=0$ and $\kappa=1$, realizes the  five dof of a massive graviton with the mass term \eqref{2.3}, constrained only by $m_1^2\geq 0$ and $m_1^2+4m_2^2\neq0$, which includes in particular  the much more restrictive Fierz-Pauli condition $m_1^2+m_2^2=0$.

So, the theory corresponding to the solution \eqref{3.11}, like all the other solutions from \eqref{3.4} to \eqref{3.10}, has well defined propagators without tachyonic poles. But, more than this,  it has been shown in \cite{Blasi:2015lrg} that it displays an unique feature with respect to the others solutions: the Fierz-Pauli case can be identified by a BRS symmetry. In this paper we see that it is remarkably peculiar for another reason: it realizes the five dof of a massive spin-2 particle without the Fierz-Pauli constraint \eqref{2.4}.

\vspace{1.5cm}

{\bf Acknowledgements}

N.M. thanks the support of INFN Scientific Initiative SFT: ``Statistical Field Theory, Low-Dimensional Systems, Integrable Models and Applications''

%*****************************************************************************

\end{document}